\documentclass[referee]{cjaa} % referee version: for submission
\usepackage{graphicx}                %for PS/EPS graphics inclusion, new
\input{epsf.sty}                     %for PS/EPS graphics inclusion, old
\input{psfig.sty}                    %for PS/EPS graphics inclusion, old

\def\ergscm2 {erg\,s$^{-1}$cm$^{-2}$}

\def\sgra{SGR\,1900+14\,}
\def\sgrb{SGR\,1806--20\,}

\begin{document}
   
\title{Magnetars' Giant Flares: the case of SGR 1806-20}
\footnotetext{$^*$ Marie Curie Fellowship; N.Rea@sron.nl}

\volnopage{Vol.0 (200x) No.0, 000--000}      %%preserved for Editor. DOn't remove!
\setcounter{page}{1}          %%starting page, preserved for Editor. DOn't remove!

\author{
Nanda Rea
\inst{1,2}\,$^*$
\and GianLuca Israel 
\inst{2}
\and Sandro Mereghetti 
\inst{3} 
\and Andrea Tiengo 
\inst{3} 
\and Silvia Zane 
\inst{4} 
\and Roberto Turolla 
\inst{5} 
\and Luigi Stella
\inst{2}
}

\offprints{N. Rea}                   %% is disabled in fact

\institute{SRON--Netherlands Institute for Space Research, Sorbonnelaan, 2, 3584 CA, Utrecht, The Netherlands \\
\and
  INAF--Astronomical Observatory of Rome, via Frascati 33, 00040, Monteporzio Catone, Rome, Italy \\
\and
  Istituto di Astrofisica Spaziale e Fisica Cosmica ``G.Occhialini'', via Bassini 15, 20133, Milano, Italy \\
\and
Mullard Space Science Laboratory, University College of London, Holbury St. Mary, Dorking Surrey, RH5 6NT, UK \\
\and
University of Padua, Physics Department, via Marzolo 8, 35131, Padova, Italy \\
}

\date{Received~~2005 month day; accepted~~2005~~month day}

\abstract{ We first review on the peculiar characteristics of the bursting and flaring activity of the Soft Gamma-ray Repeaters and Anomalous X-ray Pulsars. We then report on the properties of the SGR 1806-20's Giant Flare occurred on 2004 December 27th, with particular interest on the pre and post flare intensity/hardness correlated variability. We show that these findings are consistent with the picture of a twisted internal magnetic field which stresses the star solid crust that finally cracks, causing the giant flare (and the observed torsional oscillations). This crustal fracturing is accompanied by a simplification of the external magnetic field with a (partial) untwisting of the magnetosphere.
}

\authorrunning{Nanda Rea et al.}        

\titlerunning{Magnetars' Giant Flares: the case of SGR 1806-20}  

\maketitle

\section{Introduction}         
\label{sect:intro}
%\hspace{15pt}%                   %% preserved for Editor

Soft $\gamma$-ray Repeaters (SGRs) and Anomalous X-ray Pulsars (AXPs)
are two peculiar groups of neutron stars (NSs) which stand apart from
other known classes of X-ray sources. They are all radio-quiet,
exhibit X-ray pulsations with spin periods in the $\sim$5-12\,s
range, a large spin-down rate ($\dot{P}\approx$
10$^{-10}$-10$^{-13}$s\,s$^{-1}$) and a rather high X-ray luminosity
($L_{X}\approx$10$^{34}$-10$^{36}$erg\,s$^{-1}$; for a recent review see
Woods \& Thompson~2004). The nature of the X-ray emission from these
sources has been intriguing all along. In fact, for both AXPs and
SGRs, the X-ray luminosity is too high to be produced by rotational
energy losses alone, as for more common isolated radio pulsars.

The magnetic fields of SGRs and AXPs, as estimated from the classical
dipole braking formula B$\sim$3.2$\times$10$^{19}\sqrt{P\dot{P}}$~G, are
all above the electron critical magnetic field, $B_{QED}\sim$4.4$\times
$10$^{13}$~G. At the same time, the lack of observational signatures of
a companion strongly argues against an accretion-powered binary
system, favouring instead scenarios involving isolated NSs. These
findings led to the idea that the two classes of sources are linked
together, and their X-ray emission related to their very high magnetic
field.  At present, the model which is most successful in explaining
the peculiar observational properties of SGRs and AXPs is the
``magnetar'' model. In this scenario SGRs and AXPs are thought to be
isolated NSs endowed with ultra-high magnetic fields (B$\sim$
10$^{14}$-10$^{15}$\,G) and their steady X-ray emission powered by
magnetic field decay (Duncan \& Thompson~1992).

The unpredictable flaring activity of magnetar candidates make them
different from all other known classes of neutron stars.  From the
phenomenological point of view, the bursting/flaring events can be
roughly divided in three types.

\noindent {\bf i)} {\bf X/$\gamma$-ray short bursts}.
These are the most common and less energetic SGR flaring events. They
have short duration ($\sim$0.1-0.2\,s), thermal spectra, and peak
luminosity of $\sim$10$^{40}$-10$^{41}$\,erg\,s$^{-1}$, well above the
Eddinton luminosity limit for a standard neutron star. They are
irregular in time and can occur as single events or in a bunch. AXPs
short bursts are slightly different, less energetic and with longer
durations (Kaspi et al.~2003; Woods et al.~2005)

\noindent {\bf ii)} {\bf Intermediate flares}. The name comes from the
fact that they are intermediate both in duration and luminosity
between short bursts (i) and Giant Flares (iii). They have durations
in the range $\sim$1-60\,s and luminosity of
$\sim$10$^{41}$-10$^{43}$erg\,s$^{-1}$ . Sometimes intermediate flares
last longer than the pulsars' spin periods, and show clear modulation
at the star spin period. These events have been observed in SGRs, but
they were never revealed up to now from an AXP.

\noindent {\bf iii)} {\bf Giant flares}. These are by far the most energetic
($\sim$10$^{44}$-10$^{47}$erg\,s$^{-1}$) Galactic events currently
known, second only to Supernova explosions. Only three of these events
have been recorded in decades of monitoring of the high energy sky and
all from SGRs: SGR 0526-66 on 1979 March 5 (Mazets et al. 1979), SGR
1900+14 on 1998 August 27 (Hurley et al. 1999) and the last and more
energetic one on 2004 December 27 from SGR 1806-20. All of them are
characterised by a very luminous hard peak lasting a bit less than a
second, which decays rapidly into a soft pulsating (at the NS spin
period) tail lasting hundreds of second.

\section{SGR 1806-20 before the Giant Flare}
\label{sect:presgr1806}

SGR 1806-20\, is at the moment the most prolific SGR. It showed
several periods of bursting activity since the time of its discovery
in 1979 (Laros et al. 1986). RXTE observations led to the discovery of
pulsations (period P=7.47 s and period derivative
$\dot{P}$=8$\times$10$^{-11}$ s s$^{-1}$; Kouveliotou et al. 1998) and
were subsequently used to monitor the timing properties of the source,
such as the long term $\dot{P}$ variations and the evolution of the
pulse profile (Woods et al. 2002). The first high resolution X--ray
spectra of this source, reported to date, were obtained by BeppoSAX in
October 1998 and March 1999 (Mereghetti et al. 2000). These showed a
spectrum equally well described in the 2-10 keV range by a power law
with photon index $\Gamma$=1.95 or by a thermal bremsstrahlung with
temperature kT$_{tb}$=11 keV. Similar flux values were measured at
that time during all the observations, making believe SGR 1806-20\, a
fairly stable X-ray emitter, with luminosity of
$\sim$3$\times$10$^{35}$ (d/15 kpc)$^2$ erg s$^{-1}$ (2-10 keV) in the
period 1993--2001.

Thanks to the INTEGRAL hard X-ray imaging capabilities, persistent
emission from SGR 1806-20 was detected up to the $\gamma$-rays, having
a power law spectrum with photon index $\Gamma\sim$1.5--1.9 extending
up to 150 keV (Mereghetti et al. 2005a; Molkov et al. 2005).

Although a radio counterpart were never revealed in this source, ten
years ago, VLA observations with an arcminute spatial resolution
revealed an weak evidence for a variable jet-like structure, still
under debate (Vashist, Frail \& Kulkarni 1995; Frail, Vashist \&
Kulkarni 1997).

Recently, optical and IR studies of the environment of SGR 1806-20\,
found the source part of a cluster of massive stars of ~3.0--4.5
Myr. Assuming coevality, this age suggests that the progenitor of SGR
1806-20 had an initial mass greater than $\sim$50\,M$_{\odot}$. This
is consistent with the suggestion that SGRs are the end states of
massive progenitors and may suggest that only very massive stars
evolve into magnetars (Figer et al. 2005; Eikenberry et al. 2004;
Fuchs et al. 1999).

During the last two years SGR 1806-20 displayed a gradual increase in
the level of activity, as testified by the rate at which bursts were
emitted and by an increase of the soft and hard X-ray luminosity
(Woods et al. 2004), which culminated in 2004 December 27 with the
emission of the Giant Flare. This gradual brightening was detected by
XMM-Newton, through a series of observations, obtained from April 2003
to October 2004: while in 2003 the 2--10 keV emission was similar to
that seen in previous measurements with other satellites, the source
flux doubled in the following year (Mereghetti et al. 2005c).

%%%%%%%%%%%%%%%%%%%%%%%%%%%%%%%%%%%%%%%%%%%%%%%%%%%%%%%%%%

\begin{figure*}
\hbox{
\psfig{figure=rea_2005_01_fig01.eps,width=7cm,height=5.3cm,angle=-90}
\psfig{figure=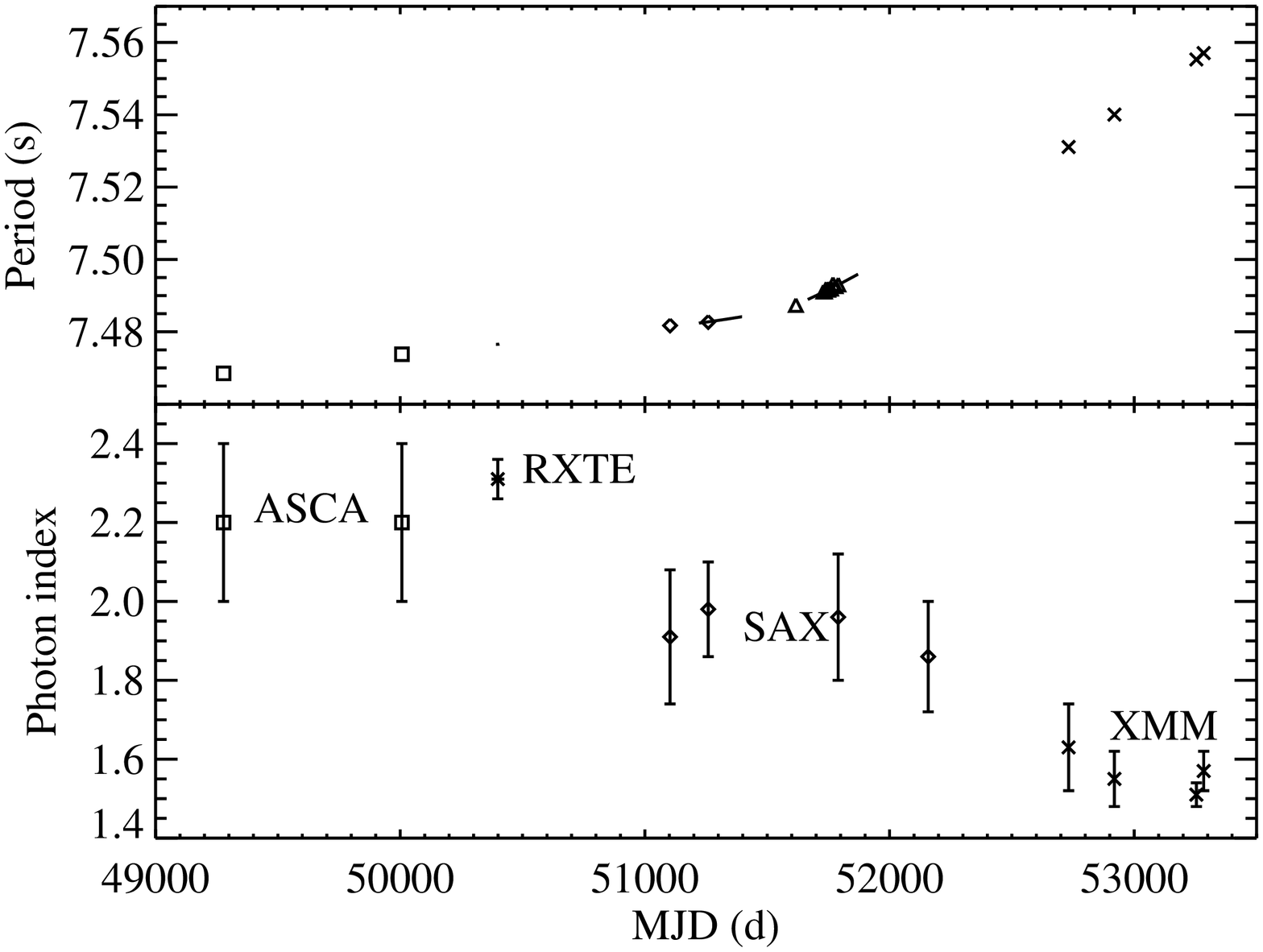,width=7cm,height=5cm,angle=90}}
\caption{{\em Left panel}: INTEGRAL light-curve of the SGR 1806-20 Giant Flare (from Mereghetti et al. 2005b); {\em Right panel}: Long term
pre flare evolution of the pulse period and power law photon
index of \sgrb\ (from Mereghetti et al. 2005c).}
\end{figure*}

%%%%%%%%%%%%%%%%%%%%%%%%%%%%%%%%%%%%%%%%%%%%%%%%%%%%%%%%%%%%%

\section{The SGR 1806-20 December 27th Giant Flare}
\label{sect:sgr1806}
%\hspace{15pt}%                   %% preserved for Editor

On 2004 December 27th, SGR 1806-20 emitted an exceptionally powerful
giant flare, with an initial hard spike lasting $\sim$0.2 s followed
by a $\sim$600\,s long pulsating tail showing about 50 cycles of
high-amplitude pulsations at the known rotation period (Hurley et
al. 2005; Palmer et al. 2005).  The prompt emission saturated almost
all $\gamma$-ray detectors, then an exact estimate of the peak fluence
was difficult to infer. The GEOTAIL (Terasawa et al. 2005) and the
SOPA (Palmer et al. 2005) geo-satellites were the only instruments not
to saturate during the peak, revealing an isotropic peak luminosity of
$\sim$2$\times$10$^{46}$\,d$^2_{15}$erg\,s$^{-1}$\, (d$^2_{15}$ is the
source distance in units of 15 kpc), a hundred time higher than that
of the two giant flares previously observed from other SGRs. The tail
energetic was instead measured by several instruments, agreeing in a
release of $\sim$5$\times$10$^{43}$\,d$^2_{15}$erg\,s$^{-1}$\,,
comparable with the other giant flares' tails.

A radio afterglow was detected (Cameron et al. 2005), with a
luminosity higher by a factor of 500 with respect to the previous
giant flare of SGR 1900+14\,, suggesting a very large difference in
the prompt burst energy. On the other hand, the consistency of the
tail energy among the three giant flares can be attributable to the
storage magnetic energy, then depending only on the source magnetic
fields, which is believed to be quite similar among the whole class of
the ``magnetars''. Interestingly, this radio afterglow emission has
been first observed as a resolved extended structure (Cameron et
al. 2005, Gaensler et al. 2005). Later on a moving structure, variable
in polarisation, was detected, which resumed the idea of a possible
jet emission from this source (Taylor et al. 2005; Fender et
al. 2005).

Very surprisingly torsional oscillations were detected during this
Giant Flare, for the first time in an isolated NS. The higher
frequency quasi periodic oscillations (QPOs) at $\sim$92.5\,Hz were
detected between 170 and 220\,s after the onset of the giant flare, in
association with an emission bump that occurred in the DC component
(and a reduction of the amplitude of the 7.56\,s pulsations). These
QPOs were detected only in the spin phase intervals away from the main
peak and reached maximum amplitude corresponding to the DC component
phase intervals. Evidence for $\sim$18 and $\sim$30\,Hz QPOs was found
between 200 and 300\,s from the onset of the giant flare, and not
obviously related to any specific interval of pulse phases (Israel et
al. 2005; see also Israel et al. in this proceedings).

\section{SGR 1806-20 after the Giant Flare}
\label{sect:postsgr1806}

After the Giant Flare event, we continued to monitor the X-ray
emission of SGR 1806-20, this time thanks to a prompt Chandra Target
of Opportunity on this source (Rea et al. 2005a). The Chandra data
clearly indicate that the spectrum softened significantly: we obtained
a power law with $\Gamma\sim$1.8. This has to be compared with the
pre-flare values $\Gamma\sim$1.2 (with the inclusion of the blackbody)
or $\Gamma\sim$1.5--1.6 (in the single power law model). The flux is
$\sim$20\% lower than the pre-flare value, but still significantly
higher than the historical flux level of $\sim$1.3$\times$10$^{-11}$
erg\,cm$^{-2}$\,s$^{-1}$. Another difference with respect to the
pre-flare properties is the smaller pulsed fraction (which changed
from about 10\% to 3\%) and the pulse profile is now double peaked.

The post-flare evolution of \sgrb\, shows both similarities and
differences when compared to that of \sgra, the only other case in
which good spectral X-ray data have been collected after a giant
flare. Even though the \sgrb\, giant flare was two orders of magnitude
more energetic than that of \sgra \ (and of SGR 0526-66 as well), it
was followed by a very rapid decay of the X-ray luminosity. We find
that the source flux has dropped below the pre-flare level after about
one month, much faster than what observed after \sgra\, giant flare.
However in both cases this flux decrease was accompanied by a
spectral softening (Woods et al. 1999; Rea et al. 2005a).  This
suggests that the post-flare softening, a feature common to both
sources, might be unrelated to the flare energetics and the decay rate
of the X-ray flux after the flare.

\section{Discussion}
\label{sect:discussion}
%\hspace{15pt}%                   %% preserved for Editor

In the recently proposed ``twisted magnetosphere'' model (Thompson,
Lyutikov \& Kulkarni 2002), it has been suggested that the internal
magnetic field of a magnetar has a strong toroidal component, which
can be comparable to the poloidal one. The presence of an internal
twist exerts a Lorentz force on the highly conducting crust
material. The net effect is to induce a rotation on the polar regions
of the crust which is contrasted by the crustal rigidity. Although the
crust may adjust quasi-plastically to the imparted stresses, from time
to time it cracks and these multiple, small-scale fractures give rise
to the shaking of the external field lines and the onset of short
bursts while bigger cracks can produce giant flares. The progressive
displacement of part of the crust produces a global twist of the
external field and currents start to flow in the
magnetosphere. Charged particles develop a large optical depth to
resonant cyclotron scattering and returning currents hit the star
surface heating it up. Both the emitted luminosity and the depth
increase with the twist angle. Thermal photons emitted by the surface
undergo resonant scattering in the magnetosphere and, since the
spectral hardness increase with depth, the steady X-ray flux is
espected to correlate with the power-law index. This implies that
following such an episode, a decrease of the flux and a softening of
the spectrum is expected.

The main observational consequences of a magnetospheric untwisting,
namely a decrease in the X-ray flux, a softening of the spectrum and a
decrease of the pulsed fraction appear to be present in the X-ray
post-flare observations.

Since this work was presented we continued to monitor the decay of the
persistent X-ray emission of SGR 1806-20 with XMM-Newton (Tiengo et
al. 2005; Rea et al. 2005b). Consistently with the twisted
magnetosphere scenario the source is still decreasing in flux and
correlatedly softening.

\label{lastpage}

\end{document}